\documentclass{osa-article}

\journal{osac}

\articletype{Research Article}

\begin{document}

\title{Design and analysis of high electron mobility transistor (HEMT) inspired III-V electro-optic modulator topologies}

\author{Pallabi Das,\authormark{1} Tian-Li Wu,\authormark{2} and Siddharth Tallur\authormark{1,*}}

\address{\authormark{1}Department of Electrical Engineering, Indian Institute of Technology (IIT) Bombay, Mumbai 400076, India\\
\authormark{2}International College of Semiconductor Technology, National Chiao Tung University, 30010 Hsinchu, Taiwan\\
}

\email{\authormark{*}stallur@ee.iitb.ac.in} 



\begin{abstract}
III-V heterostructure based high electron mobility transistors (HEMTs) offer superior performance as compared to CMOS silicon transistors owing to the high mobility in the 2D electron gas (2DEG) channel at the heterostructure interface. Gallium nitride (GaN) based HEMTs are also suitable for high power and high temperature applications. GaN has a rich offering of material properties spanning domains of nonlinear optics, piezoelectric micro-electro-mechanical systems (MEMS), and monolithic microwave integrated circuits (MMICs). In this paper, we propose HEMT inspired III-V electro-optic modulator topologies that could potentially outperform silicon photonic modulators. We analyze the electroabsorption and electrorefraction on account of the 2DEG interaction with light and present a design framework to selectively leverage the desired mechanism of modulation. Our analysis suggests that modulation index of electrorefractive modulation in a HEMT-like structure is comparable to silicon photonic modulators, albeit with much higher electron mobility and thereby much higher modulation rates.
\end{abstract}

\section{Introduction}

In the last few decades, fiber optic communication has become the technology of choice for high speed communication links with capacities of several hundreds of gigabits per second (Gbps) and low signal losses over link lengths of thousands of kilometers. While fiber optic communication forms a large part of the backbone for communication systems, data rates achievable in short-reach communication networks remain a critical bottleneck, limited by interconnects. Parasitic capacitance in electrical traces limits maximum data rates in conventional electrical interconnects to $< 10Gbps$ \cite{Miller,B2}. Silicon photonics has drawn immense interest in recent years owing to electronic-photonic integration  and CMOS-compatible processes and signal levels that it offers for optical information transfer and signal processing {\cite{lim2013review, liu2004high, heck2014ultra, B2}}. Encoding information generated by on-chip CMOS integrated circuits onto an optical carrier signal requires an electro-optic (EO) modulator. Although silicon is not an electro-optically active material, it can used as an EO modulator by exploiting carrier injection or plasma dispersion effect, when used in p-i-n (p-type/intrinsic/n-type layers) device configuration {\cite{B2, Si_plasma, zhou2006silicon}}. However the carrier recombination lifetime of free charge carriers limits the operating speeds to $\approx 50-60Gbps$ in such devices \cite{Si_40gbps}, and higher bandwidth applications typically employ wavelength division multiplexing \cite{dwivedi2015coarse}. Alternative modulation mechanisms utilizing other materials compatible with silicon technology e.g. germanium, have been demonstrated to achieve more efficient modulation\cite{Geintegrated}. Si-Ge multi-quantum well (MQW) electro-absorption modulators exploiting the quantum confined stark effect (QCSE) and Franz-Keldysh effect have also been reported\cite{Sige, sige2, B2}. QCSE has also been explored in III-V intersubband opto electronic devices \cite{nitridebook, machhadani}, despite the processing challenge it imposes in terms of incorporating the multi-quantum well structure into the device layer stack. Another alternative technology platform gaining momentum in recent years is lithium niobate (LiNbO$_{3}$)\cite{LN,LN2}. Unlike silicon, LiNbO$_{3}$ has strong EO coefficient and lithium niobate electro-optic modulators have recently been demonstrated operating at frequencies exceeding $100Gbps$ \cite{B1} albeit with inadequate bit error rates for practical applications. While most commercial optical modulator products also employ lithium niobate, the material compatibility with other platforms fundamentally limits the scope of on-chip interconnect applications of this technology.
 
Another limitation for high speed on-chip communication arises due to gain-bandwidth performance limits of transistors in any CMOS technology node. For any transistor technology node, the RF performance depends on the cut-off frequency $(f_T)$ and the maximum operating frequency of the device $(f_{max})$. Compared to the highest speed silicon CMOS transistors, gallium nitride (GaN) based High Electron Mobility Transistors (HEMTs) are emerging as a commercially viable technology for high-speed and high-power applications \cite{Gan_300ghz} owing to the rich material properties in GaN, such as large band gap, large breakdown field strength, high saturation velocity and high electron mobility \cite{nitridebook,GaN,HEMT1,hemt2}. 
A large driver of the performance superiority in GaN HEMTs is the electrically tunable, highly localized sheet of charge (two dimensional electron gas: 2DEG) formed at the III-V heterostructure interface due to spontaneous and piezoelectric polarization in the constituent materials \cite{nitridebook}.
In this paper, we propose and analyze electro-optic modulator design topologies inspired by III-V HEMTs in a GaN technology platform. Such structures display two types of electro-optic interactions mediated by the 2DEG, namely electroabsorption (due to quantum confinement of electrons in the quantum well formed at the III-V heterostructure interface), and electrorefraction (due to plasma dispersion effect of the 2DEG). We present a variety of design topologies to selectively harness the desired mechanism and compare their performance parameters. Our analysis suggests that electrorefraction is the dominant modulation mechanism due to 2DEG. Coupled with the large electron mobility in HEMT-like devices, this could be leveraged to realize high speed electrorefractive optical modulators. We present the simulation framework used in this work, and the results and observations below.


\section{Simulation setup}


Silicon EO modulators typically operate based on electrorefraction (ER) in p-i-n Si slot waveguide structures, where the electric field driven free carrier concentration can be modulated in the intrinsic region by electric field introduced by applying suitable potential difference across the $p$ and $n$ regions. Modulating the free carrier concentration results in modulation of the refractive index of the material due to plasma dispersion effect \cite{B2}. Free carrier electroabsorption and electrorefraction in bulk GaN has been studied by Soltani et al. \cite{B5}, and demonstrated in bulk GaN EO modulators by several groups \cite{EA_bulk_gan}. Electroabsorption modulator topologies based on intersubband (ISB) transition of electrons in multi-quantum well (MQW) heterostructures have been explored by several groups\cite{ISBT1,ISBT2}. However, electroabsorption and electrorefraction due to 2DEG in III-V heterostructures, and their suitability for designing EO modulators have not previously been studied.


To simulate electroabsorption due to intersubband transition, we need to solve self-consistent 1-D Schr\"{o}dinger-Poisson equation to compute the sub-band energies in the potential well formed at the heterostructure interface. This is achieved through TCAD (Technology computer-aided design) simulations performed using Silvaco ATLAS interactive tool. A simulation model of polarization charge analysis for various barrier thickness and Al concentration values in an AlGaN/GaN HEMT device has already been implemented as a built-in example in ATLAS~\cite{tcad}. We adapt this example to our device design to compute the 2DEG carrier concentration density for various values of gate voltage applied to the HEMT. 
The single mode condition for the waveguide structure is essential for any integrated opto-electronic device to sustain the fundamental mode of propagation for strong optical confinement and efficient coupling with external optical fibers without intermode interference. A hybrid numerical-analytical design methodology for obtaining single mode condition in ridge waveguides with geometries larger than the wavelength of guided light
has been presented in our previous work\cite{hybrid}. We design a ridge waveguide structure following our design methodology and perform finite difference time domain (FDTD) simulations in OptiFDTD 32-bit tool to validate the design. The optimized geometry is then simulated in Silvaco ATLAS to quantify the degree of electroabsorption and electrorefraction in the structure.

\section{Results and Discussion}

\subsection{Electrorefraction in free-space modulator topology}

\begin{figure}
\centering
\includegraphics[width=\linewidth]{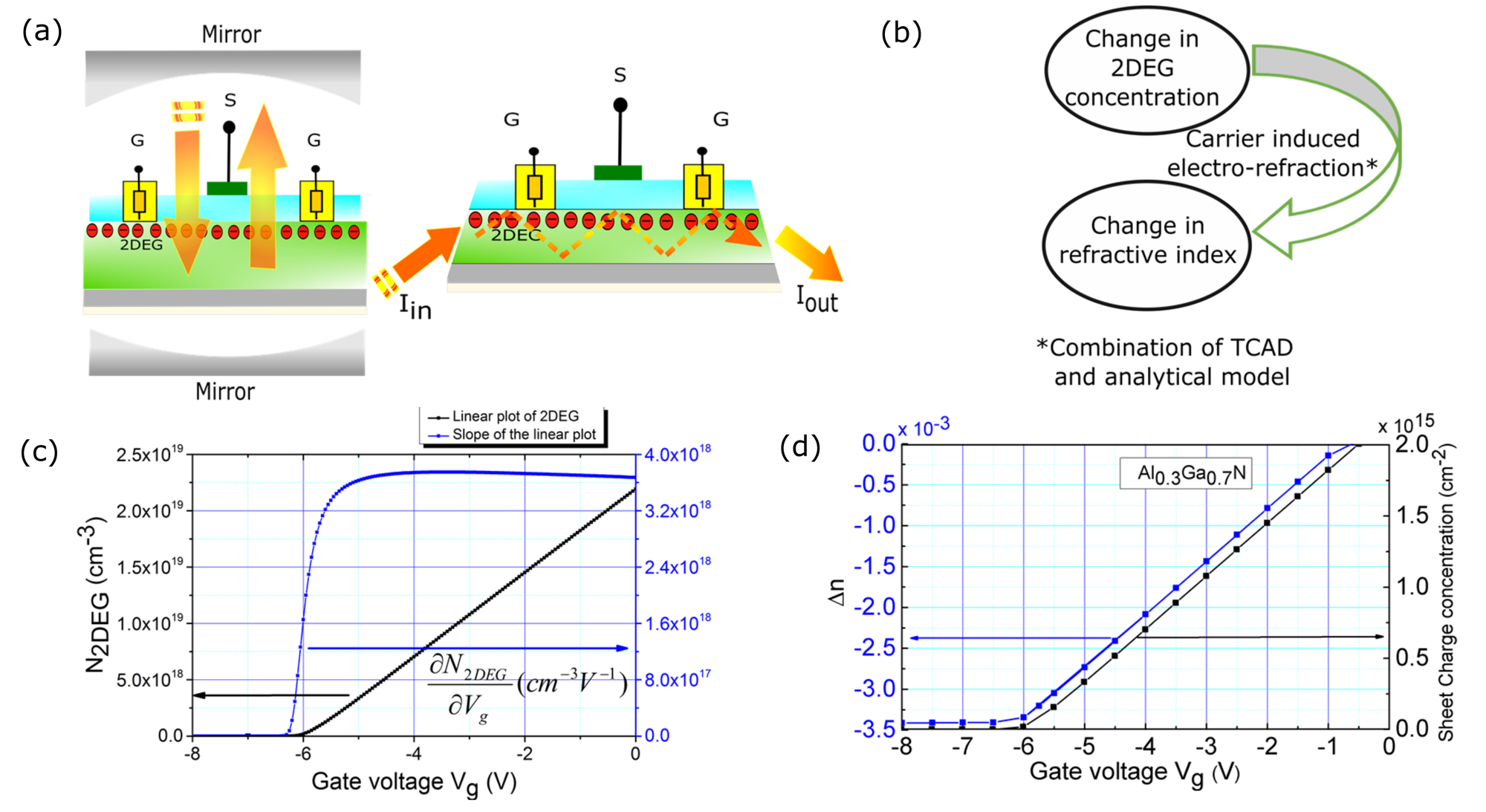}
\caption{(a) Light interaction with 2DEG in AlGaN/GaN HEMTs can be realized in a free-space manifestation (left) embedded in a Fabry Perot cavity or a multipass waveguide manifestation (right). (b) Illustration of the simulation process flow followed in this work to determine the modulation index for electrorefractive modulation due to 2DEG. (c) Variation of 2DEG charge density ($N_{2DEG}$) and slope of $(N_{2DEG})-V_g$ curve $\left(\frac{\partial N_{2DEG}}{\partial V_g}\right)$ with applied gate bias ($V_g$). (d) The variation of refractive index with applied gate voltage is well correlated to the variation in sheet charge density of the 2DEG.}
\label{fig1}
\end{figure}

The device manifestation that we consider is conceptually similar to a AlGaN/GaN HEMT where free-space light beam interacts with the 2DEG formed at the heterostructure. The light beam is incident normal to the wafer, reflects from the bottom surface of the wafer and thereby interacts with the 2DEG upon incidence as well as post reflection. The interaction can be enhanced by embedding such a device in a Fabry-Perot cavity as shown in Figure~\ref{fig1}(a). We analyze the effect of plasma dispersion effect due to the 2DEG carrier density modulated via external gate voltage applied to a depleted Schottky diode formed by the gate electrode and channel \cite{B3, B4}.
Intersubband (ISB) absorption due to TM polarized light in a quantum well is forbidden by selection rules at normal incidence \cite{ISBrules}. Computation of the absorbtion spectra due to ISB transition in a triangular potential well formed at the AlGaN/GaN interface for light incident at an angle to the normal is presented in the next section. Since there is no electric field applied across the GaN film, the linear electro-optic effect and quadratic electro-optic effect will not contribute to optical modulation in such a structure. Therefore electrorefration due to plasma dispersion effect of the 2DEG will by the sole contributor to electro-optic modulation of the free space light beam. The plasma dispersion effect results in modulation of the effective refractive index (and thereby, the phase of the carrier light beam) at the AlGaN/GaN interface due to modulation of the 2DEG carrier concentration, by varying the applied gate voltage. Applying a sufficiently large reverse bias voltage to the Schottky contact on the gate electrode completely depletes the 2DEG (OFF state of modulator), and a sufficiently large gate voltage can be applied to introduce the 2DEG (ON state of modulator).

The phase of the light beam can be modulated by an amount $(\Delta \phi)$ expressed as $\Delta \phi=\frac{2\pi}{\lambda} (2t \times \Delta n_{2DEG})$,
where $\lambda$ is the wavelength of light, $t$ is the thickness of the 2DEG, and $\Delta n_{2DEG}$ is the change in refractive index due to the 2DEG plasma dispersion effect. The expression accounts for the interaction of light with 2DEG both upon incidence and reflection, as mentioned in the previous paragraph. For an efficient modulator, a phase shift of $\pi$ radians (OFF state) should be achievable using low switching voltages. To enhance the phase shift, the device could be embedded in a photonic resonator or a high finesse Fabry-Perot cavity. The Drude plasma model captures the dependence of complex refractive index $(n + ik)$ of a material on the free carrier concentration in the semiconductor, where $n$ and $k$ denote the real and imaginary part of the refractive index respectively. The analytical expressions for the plasma dispersion effect (electrorefraction) and free-carrier absorption (electroabsorption) are given by\cite{B5} $\Delta n= \frac{-e^{2}\lambda^{2}}{8\pi^{2}c^{2}\varepsilon_{o}n_s}\bigg(\frac{\Delta N}{m^{*}_{ce}}+\frac{\Delta P}{m^{*}_{ch}}\bigg)$ and $\Delta k= \frac{e^{3}\lambda^{3}}{16\pi^{3}c^{3}\varepsilon_{o}n_s}\bigg(\frac{\Delta N}{\mu_{e} m^{*}_{ce}}+\frac{\Delta P}{\mu_{h} m^{*}_{ch}}\bigg)$ respectively, where, $\Delta N$ and $\Delta P$ are the free-carrier densities, $\mu_e$ and $\mu_h$ are the mobilities and $ m^{*}_{ce}$ and $m^{*}_{ch}$ are the conductivity effective masses for electrons and holes respectively, $e$ is the electron charge, $c$ is the speed of light, $\epsilon_o $ is the permittivity of vacuum, and $n_s$ is the real part of refractive index of the undoped semiconductor. HEMTs typically employ undoped films, and the contribution to refractive index modulation from free-carriers in the AlGaN and GaN films is therefore negligible. The refractive index modulation is thereby dominated by modulation of the 2DEG charge density on account of the gate voltage. As previously stated, the structure topology forbids electro-absorption. The modified Drude model accounting solely for electrorefraction i.e. $\Delta n_{2DEG}$ due to change in 2DEG charge density  $(\Delta N_{2DEG})$ is given by equation \eqref{del_n_modified}:
\begin{equation}
\Delta n_{2DEG}= \frac{-e^{2}\lambda^2}{8\pi^{2}c^{2}\varepsilon_{o}n_{GaN}} \bigg(\frac{\Delta N_{2DEG}}{m^{*}_{ce}}\bigg)
\label{del_n_modified}
\end{equation}

Here $n_{GaN}$ denotes the refractive index of the GaN film. Figure~\ref{fig1}(b) schematically illustrates the computational analysis flow. To validate the expression in equation \eqref{del_n_modified}, we first simulate the variation of $N_{2DEG}$ for various gate bias voltages (varied from $-8V$ to $+2V$) in Silvaco ATLAS. The source and drain region contacts are connected to ground and the DC gate voltage $(V_g)$ is swept in a range covering pinch-off (2DEG fully depleted) to enhancement (maximum 2DEG concentration). The analytical expression for $N_{2DEG}$ dependence on gate voltage $(V_g)$ has been extensively studied in literature~\cite{B7, B8, B9, dasgupta} (for details see supplementary information in appendix A), and these models are used to adjust the polarization charge factor in the simulation to reproduce the 2DEG sheet charge density $(\approx 10^{14}cm^{-2})$. Self-consistent solutions of the one-dimensional (1D)  Schr\"{o}dinger-Poisson equations are solved to evaluate the wavefunctions of the first two energy sub-bands (Figure~\ref{fig2}). The contribution of the fundamental energy level dominates the contribution of the second energy level  to $N_{2DEG}$, and therefore only the fundamental energy level contribution is considered for subsequent analysis.

We use two approaches to evaluate the dependence of $\Delta n_{2DEG}$ on $V_g$. The first approach utilizes a hybrid TCAD-analytical computation, wherein we obtain the relation of $N_{2DEG}$ to $V_g$ from TCAD simulations, which is then used to calculate the change in refractive index using equation \eqref{del_n_modified}. It is important to note that the 2DEG concentration is typically specified as a surface charge density. However, the notation used in equation (\ref{del_n_modified}) denotes the volume charge density of the 2DEG. The TCAD simulations provide 2DEG concentration in terms of volume charge density at user-specified depth from the interface. The 2DEG surface charge concentration (in $cm^{-2}$) is obtained by integrating the volume charge density (in units of $cm^{-3}$) over the thickness of the GaN film.
To evaluate the hybrid-TCAD analytical model we substitute the electron concentration (in units of $cm^{-3}$) at a depth of $1nm$ from the interface in equation \eqref{del_n_modified}. 
Figure~\ref{fig1}(c) shows the 2DEG charge density variation with gate voltage ($V_g$) for typical HEMT dimensions with $25nm$ thick Al$_{0.3}$Ga$_{0.7}$N on $300nm$ thick GaN substrate obtained with TCAD simulations performed in Silvaco ATLAS, and the derivative of the curve. The derivative $(\frac{\partial N_{2DEG}}{\partial V_g})$ of the curve signifies the sensitivity and is computed in Origin Lab. Considering the peak-peak amplitude of the modulating gate voltage to be $V_{drive} = 1V_{p-p}$ and operating wavelength $\lambda = 1.55 \mu m$, we obtain the change in volume charge density of 2DEG ($\Delta N_{2DEG}$) $\approx 3.5 \times 10^{18} cm^{-3}$ from TCAD simulations, and the corresponding change in refractive index as computed using the modified Drude model presented in equation \eqref{del_n_modified} is $\Delta n_{2DEG} \approx -7.39 \times 10^{-3}$. The values of other parameters are chosen as $n_{GaN}=2.31$ and $m^{*}_{c,2DEG}=0.22 \times m_e$ \cite{mass}, where $m_e$ is the electron rest mass $\approx 9.1 \times 10^{-31}kg$. This results in a phase shift $\Delta \phi_{2DEG} \approx -5.99 \times 10^{-5}$ radians. As mentioned earlier, the phase shift can be enhanced by embedding the device in a high finesse Fabry-Perot cavity to increase the number of round-trips of light through the 2DEG interface.

The second approach we follow relies on utilizing the built-in Drude model in Silvaco ATLAS (details of implementation of the model are presented in supplementary information in appendix B). We record the effective refractive index at the interface for various values of $V_g$. The real part of the effective refractive index as a function of applied gate voltage is plotted in Figure~\ref{fig1}(d). 
The pinch-off voltage $(V_{off})$ for complete depletion of the 2DEG is obtained from capacitance-voltage (C-V) profile simulation in Silvaco ATLAS. We observe that the knee point for the real part of the effective refractive is the same as $V_{off}$, thus cross-validating the simulations. We also observe that the 2DEG sheet charge reduces to a negligible value below $V_{off}$ ($-6V$ in this example). The maximum slope of the curve (modulation efficiency) is noticed for $V_g$ in range of $-6V$ to $-1V$. The change in effective refractive index as obtained from the TCAD simulation is $\approx -3.0 \times 10^{-3}$, which is comparable to the solution obtained from the hybrid TCAD-analytical approach. The magnitude of the change in refractive index is also comparable to conventional silicon plasma dispersion modulators~\cite{B2}, and thus the proposed architecture holds great promise for ultra-fast modulation of free-space light in a practical device realization. The limitation of the AlGaN/GaN HEMT free space modulator is the limited interaction length of light with 2DEG, which will result in diminished modulation depth. This can be improved in a waveguided implementation of this structure, where the heterostructure interface can be designed to overlap with the core of the waveguide that has the largest optical mode intensity. The interaction can be additionally enhanced in an AlGaN/GaN/AlGaN double quantum well structure. However such a structure requires significantly larger reverse bias voltages 
to deplete the 2DEG charge carriers in the well and may not be practical. Moreover the ISB absorption emerges as the dominant modulation mechanism in this structure. The detailed analysis of ISB absorption in a multi quantum well (MQW) based AlGaN/GaN electroabsorption modulator is described in the next subsection.

\subsection{Electroabsorption in guided wave modulator topology}

\begin{figure}
\centering
\includegraphics[width=\linewidth]{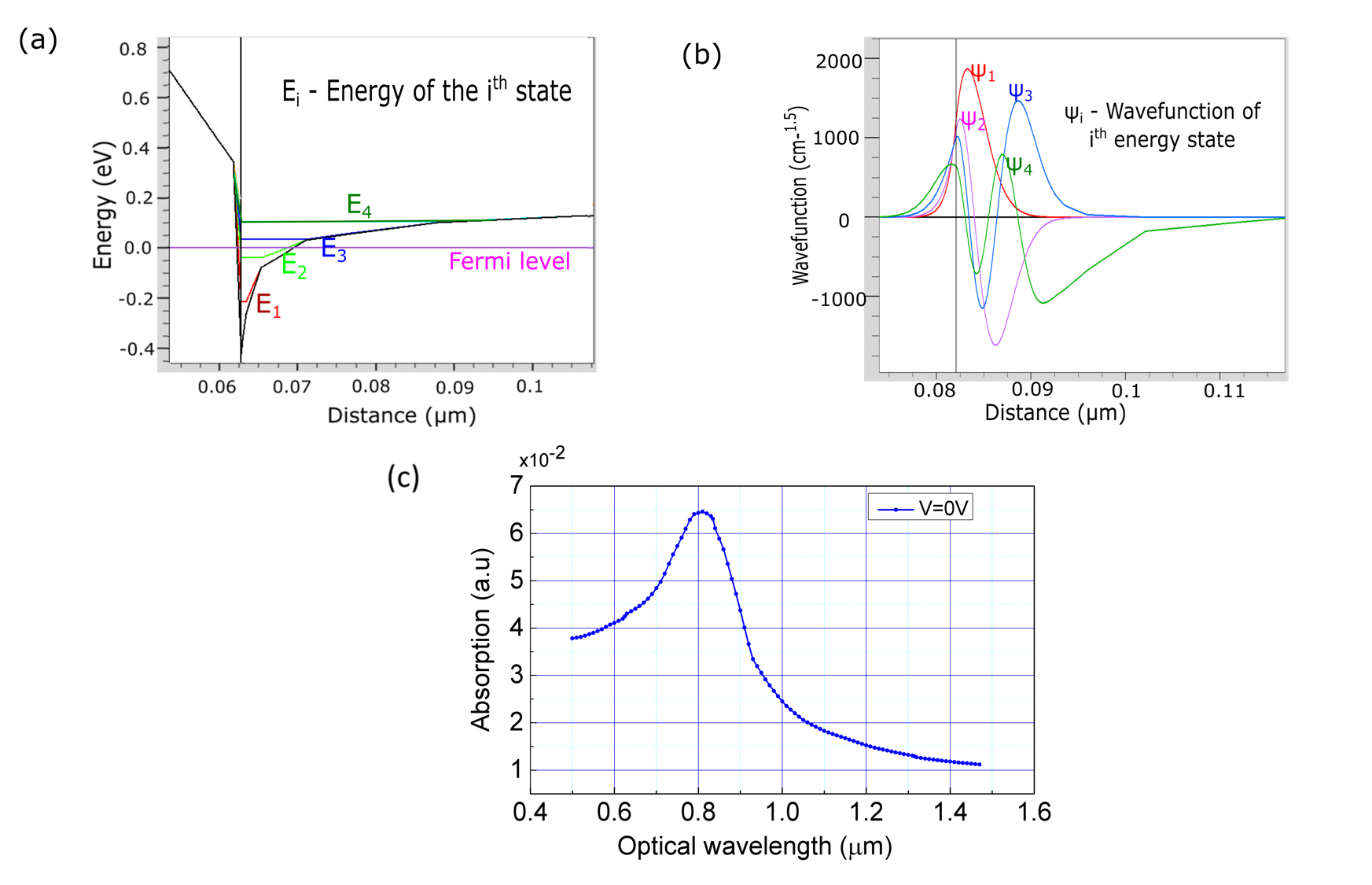}
\caption{TCAD simulation results for (a) eigen-energy levels and (b) wavefunctions for the 2DEG electrons in the triangular potential well formed at the AlGaN/GaN interface. The two lowest subbands lie below the Fermi level in the triangular potential well. (c) The absorption spectrum (at $0V$ gate voltage) obtained from TCAD simulation shows that the electroabsorption due to intersubband transitions in the triangular potential well is maximum at a wavelength of $800nm$ and negligible at $1.55 \mu m$ wavelength.}
\label{fig2}
\end{figure}

Incorporating the heterostructure into a waveguide enables enhanced overlapping of the optical mode with the 2DEG charge carriers, thereby boosting the modulation index in such a device. This can be achieved by incorporating a number of quantum wells in the waveguide core.
A variety of waveguide based optical devices based on ISB transitions in III-V quantum wells (QWs) have been reported \cite{ISBT2, AlN_GaN_ISB1, hofstetter2002midinfrared}. ISB absorption occurs only if the electric field is oriented normal to the plane of the layers comprising the quantum wells. Therefore ISB absorption due to normal incidence of TM polarized light is prohibited. Alternatively, a multi-pass ridge waveguide with wedged facets can leverage electrorefraction as well as ISB in the triangular potential well at the interface of a single AlGaN/GaN HEMT structure as shown in Figure~\ref{fig1}(a). We perform TCAD simulations to obtain the electroabsorption spectra for Al$_{0.3}$Ga$_{0.7}$N/GaN HEMT for TM polarized light incident at an angle of $30^o$ to the normal. The device geometry we consider here is an Al$_{0.3}$Ga$_{0.7}$N/GaN/GaN structure, where a $10nm$ thick undoped GaN region is defined as the quantum well, sandwiched between undoped $25nm$ thick Al$_{0.3}$Ga$_{0.7}$N and $300nm$ thick GaN layers. The energy band diagram obtained through TCAD simulation shows presence of two subbands below the Fermi level in the triangular potential well (Figure~\ref{fig2}(a)). The corresponding wavefunctions are shown in Figure~\ref{fig2}(b). The absorption spectrum obtained from the simulation (Figure~\ref{fig2}(c)) shows that the optical absorption due to ISB transition in the triangular potential well is maximum at near infrared (NIR) wavelengths in vicinity of $800nm$ and negligible for $1.5 \mu m$ wavelength light. To increase the electroabsorption at C-band and L-band wavelengths for telecommunication, we investigate the performance of multi-quantum well (MQW) structures.

\begin{figure}
\centering
\includegraphics[width=\linewidth]{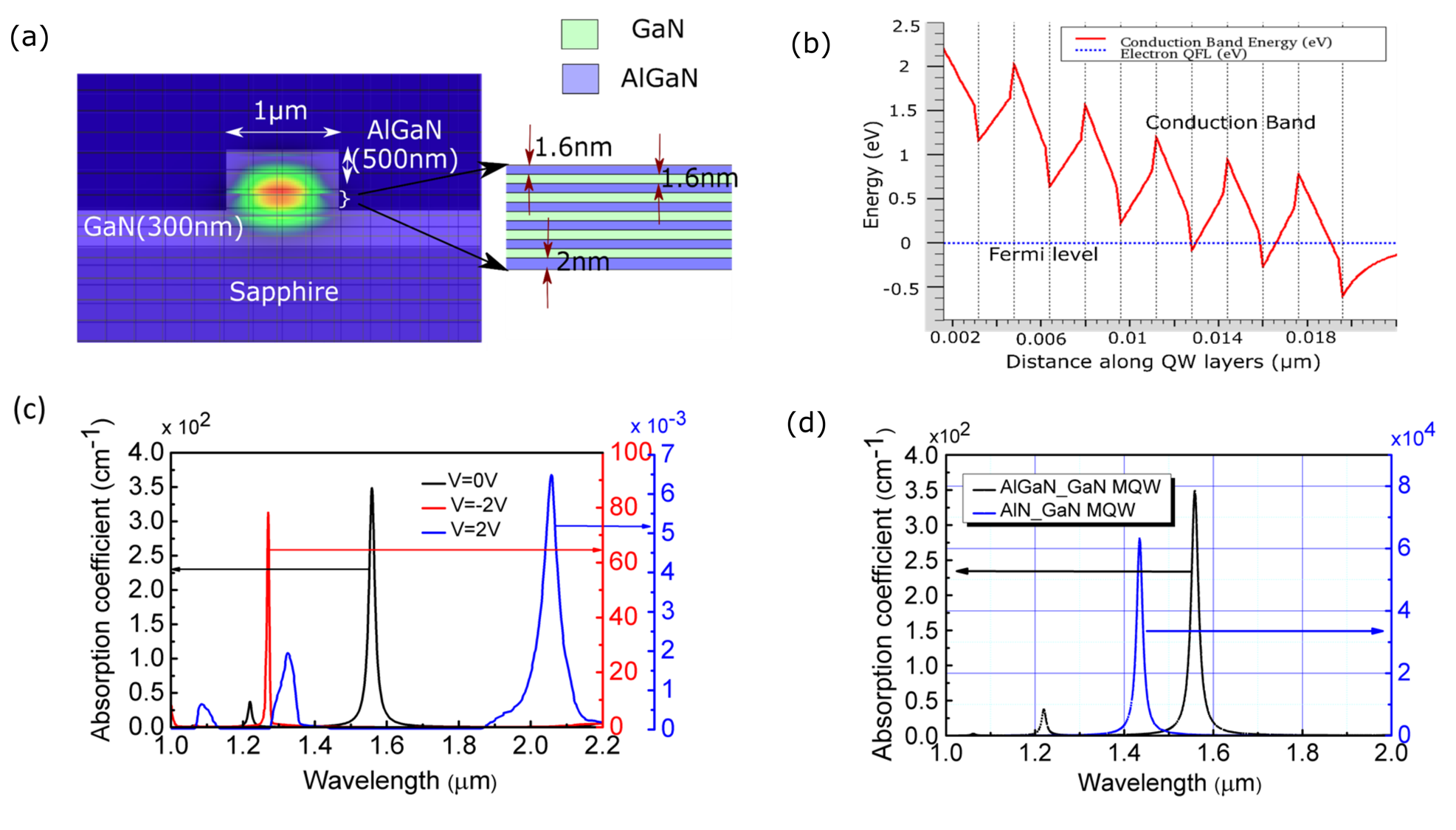}
\caption{(a) Indicative geometry of an AlGaN/GaN ridge waveguide with embedded AlGaN/GaN multi quantum wells (MQWs) designed for single mode condition. The optical mode profile simulated using OptiFDTD 32 bit software shows maximal overlap with the MQWs. (b) Energy band diagram for $5$-period Al$_{0.5}$Ga$_{0.5}$N/GaN MQWs obtained from TCAD simulation. (c) The absorption spectra obtained from TCAD simulations for the $5$-period Al$_{0.5}$Ga$_{0.5}$N/GaN MQWs for various values of gate bias voltage. The absorption peak is a strong function of the gate bias voltage and hence the MQW makes it possible to realize an electroabsorption modulator for $1.55 \mu m$ wavelength light. (d) The absorption coefficient of Al$_{0.5}$Ga$_{0.5}$N/GaN MQW structure is two orders of magnitude lower than that of a conventional AlN/GaN MQW, thus indicating that the latter is a more suitable choice for electroabsorption modulator as compared to the former. }
\label{fig3}
\end{figure}

Electroabsorption in MQW structures can also be exploited in a waveguide structure, with large overlap of the optical mode with the MQWs. Here we analyze electroabsorption due to ISB transitions in an Al$_{0.5}$Ga$_{0.5}$N/GaN MQW structure. The device structure consists of five periods of GaN wells and AlGaN barriers, with a thickness of $1.6nm$ each, sandwiched between $500nm$ thick AlGaN cladding, and terminated by $300nm$ thick GaN layer as shown in Figure~\ref{fig3}(a). We perform FDTD simulations to identify optimal geometry for a single-mode AlGaN/GaN MQW ridge waveguide. Sapphire is used as substrate in the simulation in FDTD. The width and height of the AlGaN cladding layer in the ridge geometry are $1\mu m$ and $317.6 nm$ respectively. The position of the MQWs is chosen to overlap with the region with largest light intensity in the optical mode.

To analyze the electroabsorption in the AlGaN/GaN MQW ridge waveguide, we perform TCAD simulation using Silvaco ATLAS for the geometry obtained above.
The 1-D Schr\"{o}dinger equation for an electron residing in any of the discrete energy states in a quantum well is given by $\bigg[\frac{-{\hbar}^2}{2 {m_{i}}^{\perp}}\frac{{\delta}^2}{\delta {z}^2}+V(z)\bigg]\psi_{i}(z)= {E}_{i} \psi_{i}(z)$, where ${m_{i}}^{\perp}$ is the effective mass of the electrons in the $i^{th}$ subband, and $E_i$ and $\psi_{i}$ are the corresponding energy level and wavefunction. Fermi’s golden rule is used to calculate the probability of a transition from electron from $i^{th}$ state to $f^{th}$ state in Silvaco ATLAS. The multiband spatially dependent Kronig-Penney (K-P) model is used for self consistency. The electron effective mass $(m^*)$ for both AlGaN and GaN is assumed to be $m^* = 0.22m_e $. The wells are n-doped, with dopant density $10^{19} cm^{-3}$. We study the lowest two energy states in the TCAD simulation. Piezoelectric and spontaneous polarization of the nitride semiconductor is taken into consideration which causes the band bending at the interfaces, resulting in the energy band structure shown in Figure~\ref{fig3}(b).
The gate contact is placed atop the AlGaN barrier and bottom contact is made to the bulk GaN substrate.
The simulation is performed for TM polarized light incident at an angle of $30^\circ$ to the normal. The ISB gain is computed as a function of photon energy in the MQWs in TCAD \cite{manual}.
Negative value of the gain is indicative of electroabsorption in the quantum wells. The absorption spectra obtained from TCAD simulations for the $5$-period Al$_{0.5}$Ga$_{0.5}$N/GaN quantum well structure is shown in Figure~\ref{fig3}(c). We observe an absorption peak at $1.558\mu m$ that arises due to the ISB transition of electrons from the lower subband to the upper subband in the conduction band of the bottom-most GaN QW. Upon application of a gate bias voltage at the top AlGaN layer with respect to the GaN bulk substrate, we observe a shift in the wavelength as well as magnitude of the absorption peak. Insertion of capping layer can impact the infrared absorption in MQW structures , and has been analyzed in detail by Beeler et al. in an AlN/GaN MQW structure \cite{beeler}. We have compared the absorption coefficient of Al$_{0.5}$Ga$_{0.5}$N/GaN MQW structure with that of a conventional AlN/GaN MQW with well and barrier thickness of $1.9nm$ thickness as shown in Figure~\ref{fig3}(d). We observe that the absorption coefficient for AlN/GaN is larger than the Al$_{0.5}$Ga$_{0.5}$N/GaN MQW by several orders of magnitude for the same number of quantum wells. Therefore AlN/GaN MQWs are more suitable for electroabsorption based modulator architecture, while the HEMT-like structure studied in this paper is better suited for electrorefraction based modulator presented in the previous subsection.

\section{Conclusion}
In summary, we have presented and thoroughly analyzed two electro-optic modulator topologies in an AlGaN/GaN platform, leveraging the plasma dispersion effect of the highly confined 2DEG at the heterostructure interface. A modified Drude model is presented for quantifying the plasma dispersion effect of the 2DEG and supplemented with Silvaco ATLAS based TCAD simulations. FDTD and TCAD co-simulations are presented to solve for optimal device performance. The device primarily analyzed here is a free-space electrorefractive modulator with a structure identical to an AlGaN/GaN HEMT. We have analyzed electroabsorption in the AlGaN/GaN heterostructure as well as in III-V MQW structure and observe negligible electroabsorption in an AlGaN/GaN platform as compared to conventional AlN/GaN MQWs of comparable geometry for $1.55\mu m$ wavelength light. The electrorefraction effect is found to be dominant over electroabsorption at $1.55\mu m$ wavelength in a free-space optical modulator with TM polarized light incident normal to the plane of the device. Our future work will focus on fabrication and characterization of the HEMT electrorefractive modulator presented here, including studying performance parameters that cannot be adequately captured through simulations, such as modulation depth and modulation rate limit in such devices.

\section*{Acknowledgments}
The authors thank Prof. Swaroop Ganguly and Prof. Dipankar Saha at IIT Bombay for their valuable inputs on TCAD simulations. The authors also thank Microelectronics Computation Lab (MCL) at IIT Bombay for providing access to Silvaco ATLAS. 

\section*{Disclosures}

The authors declare no competing interests.

\appendix 
\section{Analytical model and TCAD simulation of 2DEG concentration in HEMTs}

\begin{figure}[ht]
\centering
\includegraphics[width=\linewidth]{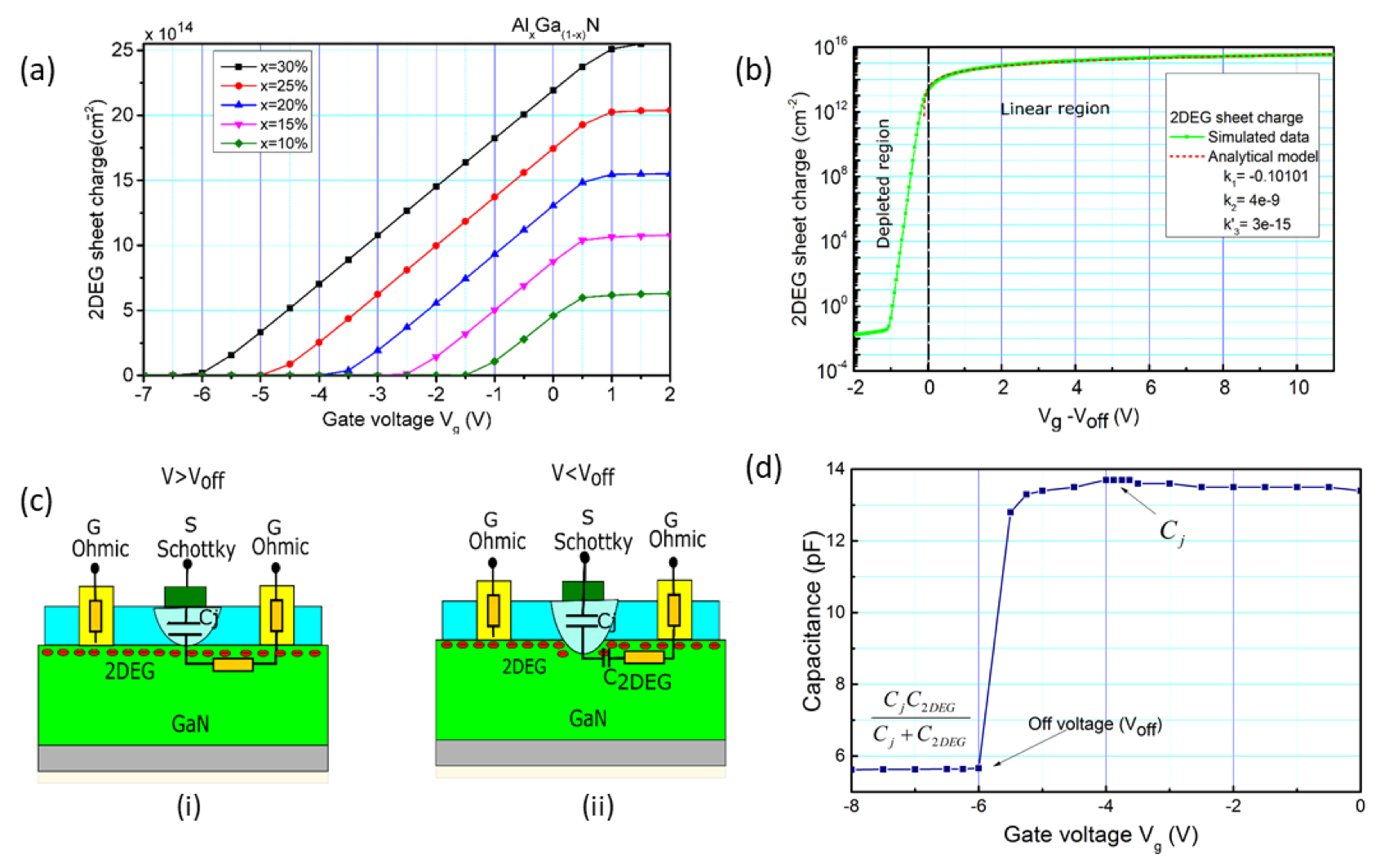}
\caption{(a) TCAD simulation showing 2DEG sheet charge concentration for various Al molar concentration $(x)$ in an Al$_x$Ga$_{1-x}$N/GaN heterostructure. (b) Variation of $N_{2DEG}$ with gate bias voltage $(V_g)$ showing comparison of simulation result (solid line) and analytical model (dotted line). Both curves show good agreement with each other. (c) (i) Junction capacitance formed by i-AlGaN region sandwiched between the Schottky metal gate and the 2DEG sheet charge; (ii) Upon application of a larger reverse bias voltage, the 2DEG is pinched off by the extension of the depletion region in to the i-GaN region, in turn producing a 2DEG capacitance that appears in series with the junction capacitance.(d) Capacitance-voltage profile of the AlGaN/GaN heterostructure obtained from TCAD simulation performed in Silvaco ATLAS.
}
\label{supp1}
\end{figure}

To derive an expression for the change in refractive index $(\Delta n_{2DEG})$ as a function of applied gate voltage $(V_g)$, we first simulate the variation of the 2DEG sheet charge density $(N_{2DEG})$ for different bias conditions in Silvaco ATLAS. The source and drain region contacts are connected to ground and the DC gate voltage is swept in a range covering pinch off to enhancement. The $N_{2DEG}$ variation with $V_g$ for different molar concentration of Al ($x$ in Al$_x$Ga$_{1-x}$N) is obtained from the simulation, as shown in Figure~\ref{supp1}(a).
For obtaining higher 2DEG concentration, we have selected $30\%$ Al molar concentration in Al$_{x}$Ga$_{1-x}$N for simulation in all subsequent discussions in this paper, unless otherwise mentioned.
The analytical expression for $N_{2DEG}$ as function of $V_g$ has been extensively studied in literature~\cite{B7, B8, khandelwal, B9, dasgupta}. These analytical models are only valid for the linear region of operation in a HEMT. Our proposed modulator relies on both depletion as well as linear regions of operation, and hence these models are not adequate. Hence we obtain the expression by solving Poisson's equation \cite{khandelwal} for our device. Assuming the AlGaN layer is completely ionized and solving Poisson's equation for metal/i-AlGaN/i-GaN layer, we obtain:
\begin{equation}
 N_{2DEG}=\frac{C_d}{q}[(V_g-V_{off})-V_{c} -E_f]
\label{ns_1}
\end{equation}
where $C_d$ is the effective capacitance per unit area between the gate electrode and 2DEG, $V_c$ is the voltage at drain/source contact and $E_f$ is the Fermi level. The term $V_{off}$ denotes the pinch-off voltage, below which the 2DEG is completely depleted. The complexity in modeling the 2DEG charge density arises due to the fact that the Fermi level $E_f$ in equation (\ref{ns_1}) is also a function of $N_{2DEG}$ and can be expressed as a transcendental equation \cite{B9}:
 \begin{equation}
N_{2DEG}= D V_{t}{ln \bigg[exp \bigg(\frac{E_f-E_0}{{kT}/q}\bigg)+1} \bigg] +ln \bigg[exp \bigg(\frac{E_f-E_1}{kT/q}\bigg)+1 \bigg]
\label{ns_2}
\end{equation}
where $D$ is the density of states, $k$ is the Boltzmann constant, $T$ is the temperature, $q$ is the charge on an electron,
and $E_0$ and $E_1$ are energies of the two lowest subbands in the triangular potential well.
In contrast to the charge control model \cite{khandelwal}, a precise analytical expression for $N_{2DEG}$ as a function of $V_g$ has been provided by Dasgupta et al. \cite{dasgupta} for AlGaAs/GaAs HEMTs. We have used similar methodology to derive the analytical model of $N_{2DEG}$ vs $V_g$ for AlGaN/GaN HEMT structure. We represent $E_f$ as a function of $N_{2DEG}$ as a polynomial expressed in equation (\ref{E_f}) \cite{dasgupta} and then employ a curve fitting technique to obtain the relation of $N_{2DEG}$ and $V_g$.
\begin{equation}
E_f= k_1 +k_{2}N_{2DEG}^{1/2}+k_3 N_{2DEG}
\label{E_f}
\end{equation}

Substituting equation~(\ref{E_f}) in equation~(\ref{ns_1}), we obtain:
\begin{equation}
N_{2DEG}=\bigg\{ {\frac{[-k_2 +[{k_2}^2 +4{k_3}^{'}(V_g -V_{off}-k_1)]^{1/2}]}{{2{k_3}^{'}}}}\bigg \}^2
\label{n_final}
\end{equation}
where 
\begin{equation}
{k_3^{'}}={k_3}+\frac{qd_{AlGaN}}{\epsilon}
\label{k3}
\end{equation}

Here $d_{AlGaN}$ is the thickness and $\epsilon$ is the permitivitty of the AlGaN layer. The values for $k_1$, $k_2$, and $k_3^{'}$ that appear in equation (\ref{n_final}) are obtained from curve fitting: $k_1= -0.10101V$ , $k_2=4 \times 10^{-9}V-cm$ , ${k_3}^{'}=3 \times 10^{-15}V-cm^2$. Figure~\ref{supp1}(b) confirms that the solution obtained from simulation and the derived analytical model show good agreement with each other.
The pinch off voltage $(V_{off})$ observed for $30\%$ Al molar concentration in Al$_{x}$Ga$_{1-x}$N from the simulation (shown in Figure~\ref{supp1}(a)) is in agreement with experimental data reported by Cordier et al. \cite{measurement}. It is to be noted that the simulation parameters are obtained for an ideal AlGaN/GaN HEMT structure with no interface traps, dislocations or any other defects. The polarization charge can be calibrated in the TCAD simulation in Silvaco ATLAS using polar.scale variable.

The value of $V_{off}$ is corroborated through capacitance-voltage (C-V) profile simulation of the heterostructure. We consider a depletion mode i-AlGaN/i-GaN HEMT structure consisting of a metal Schottky contact, and Ohmic drain and source contact. Consider the case where the Schottky contact is biased at the edge of depletion region $(V_{off})$. The thin depleted AlGaN layer between the top Schottky and the bottom 2DEG sheet charge can be considered as a parallel plate capacitance $(C_j)$, for gate voltage $V_g > V_{off}$ (as shown in Figure~\ref{supp1}(c)). This capacitance can be expressed as $C_{j}= \frac{\epsilon_r \epsilon_o A}{d_{AlGaN}}$, 
where $\epsilon_o$ and $\epsilon_r$ are the permittivity of free space and relative permittivity of the AlGaN layer respectively, $A$ is the area under the gate electrode and $d_{AlGaN}$ is thickness of the AlGaN layer. 
Upon application of a reverse bias voltage, the 2DEG is depleted and for gate voltage $V_g < V_{off}$ the heterostructure presents an additional depletion capacitance $(C_{2DEG})$. The net capacitance thus reduces at pinch-off. The total capacitance $(C_{tot})$ in the pinch off regime is given by \cite{B2,B3} $C_{tot}= \frac{C_{j}\times C_{2DEG}}{C_{j}+C_{2DEG}}$.
The modulator is turned off by reducing the gate voltage below $V_{off}$, and is turned on by increasing the gate voltage sufficiently above $V_{off}$ to have large 2DEG concentration in the channel.
The TCAD simulation of C-V profile is setup by applying an ac voltage of frequency 1MHz in Silvaco ATLAS, and the parameters used for the simulation are listed in Table 1. The simulations are carried out for AlGaN/GaN HEMT with Al molar concentration of $30\%$ and a barrier thickness of $25nm$. Figure~\ref{supp1}(d) shows the simulated C-V profile of the device. The value of $V_{off}$ for the device parameters specified in Table 1 is obtained from simulation to be $-6V$, which can be feasibly achieved in a practical implementation.

\begin{table}[!ht]
\begin{center}
\begin{tabular}{ |p{2.5cm}||p{2.5cm}|p{2.5cm}|p{2.5cm}|  }
 \hline 
 \multicolumn{4}{|c|}{Geometry parameters used for electrical simulation} \\
 \hline \hline
 Parameters & Gate electrode &Barrier layer &Channel layer\\
 \hline
 Material   & Schottky metal    &$Al_{0.3}Ga_{0.7}N$&   GaN\\
 Thickness&   500nm  & 25nm   &1.475 $\mu m$\\
 Length &4$\mu m$ & 10$\mu m$&  10 $\mu m$ \\
 \hline
\end{tabular}
\label{tab1}
\caption{Parameters used for TCAD simulation in Silvaco ATLAS to obtain C-V profile of the heterostructure and $N_{2DEG}$ variation with $V_g$.}
\end{center}
\end{table}

\section{Electro-optic TCAD simulation setup in Silvaco ATLAS}

TCAD simulation for Si optical modulator in a p-i-n waveguide configuration is provided as a built-in example in ATLAS \cite{Si_pin}. The example model implements free carrier plasma dispersion effect using ABS.FCARRIER model. The model also simulates the change in effective refractive index with applied bias voltage using WCGD.REFR parameter and WAVEGUIDE statement to define geometry and physical models for a stand-alone optical mode simulation. The parameters of the free carrier Plasma dispersion model can be modified for different materials using FC.AN, FC.AP, FC.RN, FC.RP etc. parameters in the MATERIAL section. For detailed explanation of the terms and glossary, readers are encouraged to read the Atlas User's Manual \cite{manual}. Drude plasma model based on the free carrier plasma dispersion effect has also been reported in literature for simulation of an optical modulator with a MOS junction using Silvaco Atlas \cite{tcad_drude}, where free carrier elecrto-refraction $(\Delta n)$ and electro-absorption $(\Delta k)$ models have been implemented using ABS.DRUDE parameter. We adapt this model for our device simulation. Various material properties such as electron effective mass, mobility and refractive index for different materials can be overridden manually by using DRUDE.ME, DRUDE.MUE, REAL.INDEX and IMAG.INDEX respectively in the MATERIAL statement. The parameters and their values used for this simulation are listed in Table 2.
\begin{table}[!ht]
\begin{center}
\renewcommand\thetable{2}
\begin{tabular}{ |p{3.5cm}||p{2.5cm}|p{2cm}|p{4.5cm}| }
 \hline 
 Description & Parameter & Value & Remarks \\
 \hline
 Electron effective mass   & DRUDE.ME    &  $0.22$ & $m^{*}_{c,2DEG}=0.22 \times m_e$ \\
 Electron mobility & DRUDE.MUE  & $1000 m^2/(V-s)$ & The value of the conduction band electron mobility is overridden by 2DEG mobility in Drude model \\
 Refractive index of GaN &REAL.INDEX & $2.31$ & For wavelength  $\lambda = 1.55 \mu m$  \\
 Refractive index of Al$_{x}$Ga$_{1-x}$N & REAL.INDEX & $2.15$ & For wavelength  $\lambda = 1.55 \mu m$  and $x=0.3$\\
 Polarization charge &polar.scale & $1.1$ & To modify the sign and magnitude of polarization charge (default $=1$). \\
 Numerical method &Newton TRAP & Maxtrap $=20$ & Maxtrap specifies the number of times the trap (bias step) will be repeated in case of divergence (default $=4$) \\
 Complex Eigen value solver &index.model & 1& Specifies whether the simple refractive index model (INDEX.MODEL $=0$) or the complex refractive index (INDEX.MODEL $=1$) is used \\
 \hline
\end{tabular}
 \label{tab2}
\end{center}
  \caption{Parameters used for optical TCAD simulation in Silvaco ATLAS. A detailed glossary is available in the user manual \cite{manual}.}
  \end{table}

We use POLARIZATION and CALC.STRAIN in the MODEL statement to specify the piezoelectric and spontaneous polarization respectively. POLAR.SCALE is the polarization scale factor (with a default value $=1$), whose value is set to $1.1$ in order to obtain polarization charge of the order of $\approx 10^{14} cm^{-2}$. For simulating the complex refractive index, we set the parameter INDEX.MODEL $=1$, and numerical analysis is performed using `Newton TRAP' method. We use the PROBE statement to record the effective refractive index at the interface for each value of the gate bias voltage.

\bibliography{ref}

\begin{thebibliography}{10}
\newcommand{\enquote}[1]{``#1''}

\bibitem{Miller}
D.~A.~B. Miller, \enquote{{Optical interconnects to electronic chips},}
  {\protect\JournalTitle{Appl. Opt.}} \textbf{49}, F59--F70 (2010).

\bibitem{B2}
G.~T. Reed, G.~Mashanovich, F.~Y. Gardes, and D.~J. Thomson, \enquote{{Silicon
  optical modulators},} {\protect\JournalTitle{Nature Photonics}} \textbf{4},
  518 (2010).

\bibitem{lim2013review}
A.~E.-J. Lim, J.~Song, Q.~Fang, C.~Li, X.~Tu, N.~Duan, K.~K. Chen, R.~P.-C.
  Tern, and T.-Y. Liow, \enquote{{Review of silicon photonics foundry
  efforts},} {\protect\JournalTitle{IEEE Journal of Selected Topics in Quantum
  Electronics}} \textbf{20}, 405--416 (2013).

\bibitem{liu2004high}
A.~Liu, R.~Jones, L.~Liao, D.~Samara-Rubio, D.~Rubin, O.~Cohen, R.~Nicolaescu,
  and M.~Paniccia, \enquote{{A high-speed silicon optical modulator based on a
  metal--oxide--semiconductor capacitor},} {\protect\JournalTitle{Nature}}
  \textbf{427}, 615 (2004).

\bibitem{heck2014ultra}
M.~J. Heck, J.~F. Bauters, M.~L. Davenport, D.~T. Spencer, and J.~E. Bowers,
  \enquote{{Ultra-low loss waveguide platform and its integration with silicon
  photonics},} {\protect\JournalTitle{Laser \& Photonics Reviews}} \textbf{8},
  667--686 (2014).

\bibitem{Si_plasma}
F.~Gardes, G.~Mashanovich, and G.~Reed, \enquote{{Evolution of optical
  modulation in silicon-on-insulator devices},} {\protect\JournalTitle{SPIE
  Newsroom, Dec}} \textbf{27} (2007).

\bibitem{zhou2006silicon}
L.~Zhou and A.~W. Poon, \enquote{{Silicon electro-optic modulators using pin
  diodes embedded 10-micron-diameter microdisk resonators},}
  {\protect\JournalTitle{Optics express}} \textbf{14}, 6851--6857 (2006).

\bibitem{Si_40gbps}
L.~Liao, A.~Liu, D.~Rubin, J.~Basak, Y.~Chetrit, H.~Nguyen, R.~Cohen,
  N.~Izhaky, and M.~Paniccia, \enquote{{40 Gbit/s silicon optical modulator for
  high-speed applications},} {\protect\JournalTitle{Electronics letters}}
  \textbf{43}, 1196--1197 (2007).

\bibitem{dwivedi2015coarse}
S.~Dwivedi, P.~De~Heyn, P.~Absil, J.~Van~Campenhout, and W.~Bogaerts,
  \enquote{{Coarse wavelength division multiplexer on silicon-on-insulator for
  100 GbE},} in \emph{2015 IEEE 12th International Conference on Group IV
  Photonics (GFP),}  (IEEE, 2015), pp. 9--10.

\bibitem{Geintegrated}
P.~Chaisakul, D.~Marris-Morini, J.~Frigerio, D.~Chrastina, M.-S. Rouifed,
  S.~Cecchi, P.~Crozat, G.~Isella, and L.~Vivien, \enquote{{Integrated
  germanium optical interconnects on silicon substrates},}
  {\protect\JournalTitle{Nature Photonics}} \textbf{8}, 482 (2014).

\bibitem{Sige}
Y.-H. Kuo, Y.~K. Lee, Y.~Ge, S.~Ren, J.~E. Roth, T.~I. Kamins, D.~A. Miller,
  and J.~S. Harris, \enquote{{Strong quantum-confined Stark effect in germanium
  quantum-well structures on silicon},} {\protect\JournalTitle{Nature}}
  \textbf{437}, 1334 (2005).

\bibitem{sige2}
Y.~Rong, Y.~Ge, Y.~Huo, M.~Fiorentino, M.~R. Tan, T.~I. Kamins, T.~J. Ochalski,
  G.~Huyet, and J.~S. Harris~Jr, \enquote{{Quantum-confined Stark effect in
  Ge/SiGe quantum wells on Si},} {\protect\JournalTitle{IEEE Journal of
  selected topics in quantum electronics}} \textbf{16}, 85--92 (2009).

\bibitem{nitridebook}
D.~J. Piprek, ed., \emph{{Nitride Semiconductor Devices: Principles and
  Simulation}} (John Wiley $\&$ Sons, Ltd, 2007).

\bibitem{machhadani}
H.~Machhadani, P.~Kandaswamy, S.~Sakr, A.~Vardi, A.~Wirtm{\"u}ller, L.~Nevou,
  F.~Guillot, G.~Pozzovivo, M.~Tchernycheva, A.~Lupu \emph{et~al.},
  \enquote{{GaN/AlGaN intersubband optoelectronic devices},}
  {\protect\JournalTitle{New Journal of Physics}} \textbf{11}, 125023 (2009).

\bibitem{LN}
M.~He, M.~Xu, Y.~Ren, J.~Jian, Z.~Ruan, Y.~Xu, S.~Gao, S.~Sun, X.~Wen, L.~Zhou,
  L.~Liu, C.~Guo, H.~Chen, S.~Yu, L.~Liu, and X.~Cai, \enquote{High-performance
  hybrid silicon and lithium niobate mach–zehnder modulators for 100 gbps and
  beyond,} {\protect\JournalTitle{Nature Photonics}} \textbf{13}, 1 (2019).

\bibitem{LN2}
L.~Chen, Q.~Xu, M.~G. Wood, and R.~M. Reano, \enquote{Hybrid silicon and
  lithium niobate electro-optical ring modulator,}
  {\protect\JournalTitle{Optica}} \textbf{1}, 112--118 (2014).

\bibitem{B1}
C.~Wang, M.~Zhang, B.~Stern, M.~Lipson, and M.~Lon\v{c}ar,
  \enquote{{Nanophotonic lithium niobate electro-optic modulators},}
  {\protect\JournalTitle{Opt. Express}} \textbf{26}, 1547--1555 (2018).

\bibitem{Gan_300ghz}
J.~W. {Chung}, W.~E. {Hoke}, E.~M. {Chumbes}, and T.~{Palacios},
  \enquote{{AlGaN/GaN HEMT With 300-GHz $f_{\max}$},}
  {\protect\JournalTitle{IEEE Electron Device Letters}} \textbf{31}, 195--197
  (2010).

\bibitem{GaN}
M.~Shur, \enquote{{GaN based transistors for high power applications},}
  {\protect\JournalTitle{Solid-State Electronics}} \textbf{42}, 2131--2138
  (1998).

\bibitem{HEMT1}
X.~{Huang}, Z.~{Liu}, Q.~{Li}, and F.~C. {Lee}, \enquote{Evaluation and
  application of 600 v gan hemt in cascode structure,}
  {\protect\JournalTitle{IEEE Transactions on Power Electronics}} \textbf{29},
  2453--2461 (2014).

\bibitem{hemt2}
C.~Yang, X.~Luo, T.~Sun, A.~Zhang, D.~Ouyang, S.~Deng, J.~Wei, and B.~Zhang,
  \enquote{High breakdown voltage and low dynamic on-resistance algan/gan hemt
  with fluorine ion implantation in sin x passivation layer,}
  {\protect\JournalTitle{Nanoscale research letters}} \textbf{14}, 191 (2019).

\bibitem{B5}
M.~Soltani and R.~Soref, \enquote{{Free-carrier electrorefraction and
  electroabsorption in wurtzite GaN},} {\protect\JournalTitle{Opt. Express}}
  \textbf{23}, 24984--24990 (2015).

\bibitem{EA_bulk_gan}
C.-K. Kao, A.~Bhattacharyya, C.~Thomidis, R.~Paiella, and T.~D. Moustakas,
  \enquote{Electroabsorption modulators based on bulk gan films and gan/algan
  multiple quantum wells,} {\protect\JournalTitle{Journal of Applied Physics}}
  \textbf{109}, 083102 (2011).

\bibitem{ISBT1}
J.~Heber, C.~Gmachl, H.~Ng, and A.~Cho, \enquote{{Comparative study of
  ultrafast intersubband electron scattering times at~ 1.55 $\mu$ m wavelength
  in GaN/AlGaN heterostructures},} {\protect\JournalTitle{Applied physics
  letters}} \textbf{81}, 1237--1239 (2002).

\bibitem{ISBT2}
H.~Machhadani, P.~Kandaswamy, S.~Sakr, A.~Vardi, A.~Wirtm{\"u}ller, L.~Nevou,
  F.~Guillot, G.~Pozzovivo, M.~Tchernycheva, A.~Lupu \emph{et~al.},
  \enquote{{GaN/AlGaN intersubband optoelectronic devices},}
  {\protect\JournalTitle{New Journal of Physics}} \textbf{11}, 125023 (2009).

\bibitem{tcad}
\enquote{{{Polarization Charge Analysis: TCAD} example},}
  URL{https://www.silvaco.com/examples/tcad/section20/example18\\/index.html}.

\bibitem{hybrid}
P.~Roy, P.~Das, and S.~Tallur, \enquote{{Hybrid Numerical-Analytical Effective
  Index Method for Designing Large Geometry Ridge Waveguides},} in \emph{2018
  IEEE Photonics Conference (IPC),}  (IEEE, 2018), pp. 1--2.

\bibitem{B3}
A.~{Ansari} and M.~{Rais-Zadeh}, \enquote{{A Thickness-Mode {AlGaN/GaN}
  Resonant Body High Electron Mobility Transistor},}
  {\protect\JournalTitle{IEEE Transactions on Electron Devices}} \textbf{61},
  1006--1013 (2014).

\bibitem{B4}
A.~Ansari and M.~Rais-Zadeh, \enquote{{Depletion-mediated piezoelectric
  {AlGaN/GaN} resonators},} {\protect\JournalTitle{Physica Status Solidi (A)}}
  \textbf{213}, 3007--3013 (2016).

\bibitem{ISBrules}
R.~Yang, J.~Xu, and M.~Sweeny, \enquote{Selection rules of intersubband
  transitions in conduction-band quantum wells,}
  {\protect\JournalTitle{Physical review. B, Condensed matter}} \textbf{50},
  7474--7482 (1994).

\bibitem{B7}
O.~Ambacher, J.~Smart, J.~Shealy, N.~Weimann, K.~Chu, M.~Murphy, W.~Schaff,
  L.~Eastman, R.~Dimitrov, L.~Wittmer \emph{et~al.}, \enquote{{Two-dimensional
  electron gases induced by spontaneous and piezoelectric polarization charges
  in N-and Ga-face AlGaN/GaN heterostructures},} {\protect\JournalTitle{Journal
  of applied physics}} \textbf{85}, 3222--3233 (1999).

\bibitem{B8}
J.~Zhang, B.~Syamal, X.~Zhou, S.~Arulkumaran, and G.~I. Ng, \enquote{{A compact
  model for generic MIS-HEMTs based on the unified 2DEG density expression},}
  {\protect\JournalTitle{IEEE Transactions on Electron Devices}} \textbf{61},
  314--323 (2014).

\bibitem{B9}
S.~Khandelwal, N.~Goyal, and T.~A. Fjeldly, \enquote{{A physics-based
  analytical model for 2DEG charge density in AlGaN/GaN HEMT devices},}
  {\protect\JournalTitle{IEEE Transactions on Electron Devices}} \textbf{58},
  3622--3625 (2011).

\bibitem{dasgupta}
N.~DasGupta and A.~DasGupta, \enquote{{An analytical expression for sheet
  carrier concentration vs gate voltage for HEMT modelling},}
  {\protect\JournalTitle{Solid-state electronics}} \textbf{36}, 201--203
  (1993).

\bibitem{mass}
A.~Kurakin, S.~Vitusevich, S.~Danylyuk, H.~Hardtdegen, N.~Klein, Z.~Bougrioua,
  A.~Naumov, and A.~Belyaev, \enquote{{Quantum confinement effect on the
  effective mass in two-dimensional electron gas of AlGaN/GaN
  heterostructures},} {\protect\JournalTitle{Journal of applied physics}}
  \textbf{105}, 073703 (2009).

\bibitem{AlN_GaN_ISB1}
E.~Baumann, F.~R. Giorgetta, D.~Hofstetter, S.~Leconte, F.~Guillot,
  E.~Bellet-Amalric, and E.~Monroy, \enquote{{Electrically adjustable
  intersubband absorption of a GaN/AlN superlattice grown on a transistorlike
  structure},} {\protect\JournalTitle{Applied physics letters}} \textbf{89},
  101121 (2006).

\bibitem{hofstetter2002midinfrared}
D.~Hofstetter, L.~Diehl, J.~Faist, W.~J. Schaff, J.~Hwang, L.~F. Eastman, and
  C.~Zellweger, \enquote{{Midinfrared intersubband absorption on
  AlGaN/GaN-based high-electron-mobility transistors},}
  {\protect\JournalTitle{Applied physics letters}} \textbf{80}, 2991--2993
  (2002).

\bibitem{manual}
\enquote{{{Atlas Users Manual: TCAD} publications},} .

\bibitem{beeler}
M.~Beeler, E.~Trichas, and E.~Monroy, \enquote{{III-nitride semiconductors for
  intersubband optoelectronics:a review},} {\protect\JournalTitle{Semiconductor
  Science and Technology}} \textbf{28}, 074022 (2013).

\bibitem{khandelwal}
S.~Khandelwal, N.~Goyal, and T.~A. Fjeldly, \enquote{{A physics-based
  analytical model for 2DEG charge density in AlGaN/GaN HEMT devices},}
  {\protect\JournalTitle{IEEE Transactions on Electron Devices}} \textbf{58},
  3622--3625 (2011).

\bibitem{measurement}
Y.~Cordier, J.-C. Moreno, N.~Baron, E.~Frayssinet, S.~Chenot, B.~Damilano, and
  F.~Semond, \enquote{Demonstration of algan/gan high-electron-mobility
  transistors grown by molecular beam epitaxy on si (110),}
  {\protect\JournalTitle{IEEE Electron Device Letters}} \textbf{29}, 1187--1189
  (2008).

\bibitem{Si_pin}
\enquote{{Silicon p-i-n Waveguide Modulator: TCAD example},}
  URL{https://www.silvaco.com/examples/tcad/section26/example12/index.html}.

\bibitem{tcad_drude}
\enquote{{Simulation of Si Optical Modulator with a MOS Junction: TCAD
  publications},}
  URL{https://www.silvaco.com/tech\_lib\_TCAD/simulationstandard/2015/jan\_feb\_mar/a1/a1.html}.

\end{thebibliography}






\end{document}